\documentclass[12pt,twoside]{article}
\usepackage{amssymb}

\usepackage{amsmath}
\usepackage{theorem}
\def\1{{1\mskip-10mu1}}

\def\bea{\begin{eqnarray*}}
\def\eea{\end{eqnarray*}}
\def\bean{\begin{eqnarray}}
\def\eean{\end{eqnarray}}

\newtheorem{ftheo}{THEOREM}[section]
\newtheorem{fex}[ftheo]{EXAMPLE}
\newtheorem{fdef}[ftheo]{DEFINITION}

\newtheorem{flemma}[ftheo]{LEMMA}

\newtheorem{frem}[ftheo]{REMARK}

\begin{document}

\author{Karl-Heinz Fichtner \\
Friedrich-Schiller-Universit\"{a}t Jena \\
Fakult\"{a}t f\"{u}r Mathematik und Informatik \\
Institut f\"{u}r Angewandte Mathematik \\
D-07740 Jena Deutschland \\
E-mail: fichtner@minet.uni-jena.de\\
and\\
Masanori Ohya\\
Department of Information Sciences\\
Science University of Tokyo\\
Chiba 278-8510 Japan\\
E-mail: ohya@is.noda.sut.ac.jp}
\title{Quantum Teleportation with Entangled States\\
given by Beam Splittings}
\date{August 1999}
\maketitle

\begin{abstract}
Quantum teleportation is rigorously discussed with coherent entangled states
given by beam splittings. The mathematical scheme of beam splitting has been
used to study quantum communication \cite{AO2} and quantum stochastic \cite
{FFL}. We discuss the teleportation process by means of coherent states in
this scheme for the following two cases: (1) Delete the vacuum part from
coherent states, whose compensation provides us a perfect teleportation from
Alice to Bob. (2) Use fully realistic (physical) coherent states, which
gives a non-perfect teleportation but shows that it is exact when the
average energy (density) of the coherent vectors goes to infinity.
\end{abstract}

\pagestyle{myheadings}
\markboth{\hfill FICHTNER AND OHYA \hfill}
{\hfill TELEPORTATION AND ENTANGLED STATES \hfill}  It is the paper
\cite{Ben} that the quantum teleportation was first studied as a part of
quantum cryptolgraphy \cite{Eke}. This teleportation scheme can be
mathematically expressed in the following steps \cite{IOS}:

\begin{description}
\item[Step 0:]  A girl named Alice has an unknown quantum state $\rho $ on
(a $N$--dimensional) Hilbert space $\mathcal{H}_{1}$ and she was asked to
teleport it to a boy named Bob.

\item[Step 1:]  For this purpose, we need two other Hilbert spaces $\mathcal{%
H}_{2}$ and $\mathcal{H}_{3}$, $\mathcal{H}_{2}$ is attached to Alice and $%
\mathcal{H}_{3}$ is attached to Bob. Prearrange a so-called entangled state $%
\sigma $ on $\mathcal{H}_{2}\otimes \mathcal{H}_{3}$ having certain
correlations and prepare an ensemble of the combined system in the state $%
\rho \otimes \sigma $ on $\mathcal{H}_{1}\otimes \mathcal{H}_{2}\otimes
\mathcal{H}_{3}$.

\item[Step 2:]  One then fixes a family of mutually orthogonal projections $%
(F_{nm})_{n,m=1}^{N}$ on the Hilbert space $\mathcal{H}_{1}\otimes \mathcal{H%
}_{2}$ corresponding to an observable $F:=\sum\limits_{n,m}z_{n,m}F_{nm}$,
and for a fixed one pair of indices $n,m$, Alice performs a first kind
incomplete measurement, involving only the $\mathcal{H}_{1}\otimes \mathcal{H%
}_{2}$ part of the system in the state $\rho \otimes \sigma $, which filters
the value $z_{nm}$, that is, after measurement on the given ensemble $\rho
\otimes \sigma $ of identically prepared systems, only those where $F$ shows
the value $z_{nm}$ are allowed to pass. According to the von Neumann rule,
after Alice's measurement, the state becomes
\[
\rho _{nm}^{(123)}:=\frac{(F_{nm}\otimes \mathbf{1})\rho \otimes \sigma
(F_{nm}\otimes \mathbf{1})}{\mathrm{tr}_{123}(F_{nm}\otimes \mathbf{1})\rho
\otimes \sigma (F_{nm}\otimes \mathbf{1})}
\]
where $\mathrm{tr}_{123}$ is the full trace on the Hilbert space $\mathcal{H}%
_{1}\otimes \mathcal{H}_{2}\otimes \mathcal{H}_{3}$.

\item[Step 3:]  Bob is informed which measurement was done by Alice. This is
equivalent to transmit the information that the eigenvalue $z_{nm}$ was
detected. This information is transmitted from Alice to Bob without
disturbance and by means of classical tools.

\item[Step 4:]  Making only partial measurements on the third part on the
system in the state $\rho _{nm}^{(123)}$ means that Bob will control a state
$\Lambda _{nm}(\rho )$ on $\mathcal{H}_{3}$ given by the partial trace on $%
\mathcal{H}_{1}\otimes \mathcal{H}_{2}$ of the state $\rho _{nm}^{(123)}$
(after Alice's measurement)
\begin{eqnarray*}
\Lambda _{nm}(\rho ) &=&\mathrm{tr}_{12}\;\rho _{nm}^{(123)} \\
&=&\mathrm{tr}_{12}\frac{(F_{nm}\otimes \mathbf{1})\rho \otimes \sigma
(F_{nm}\otimes \mathbf{1})}{\mathrm{tr}_{123}(F_{nm}\otimes \mathbf{1})\rho
\otimes \sigma (F_{nm}\otimes \mathbf{1)}}
\end{eqnarray*}
Thus the whole teleportation scheme given by the family $(F_{nm})$ and the e
ntangled state $\sigma $ can be characterized by the family $(\Lambda _{nm})$
of channels from the set of states on $\mathcal{H}_{1}$ into the set of
states on $\mathcal{H}_{3}$ and the family $(p_{nm})$ given by
\[
p_{nm}(\rho ):=\mathrm{tr}_{123}(F_{nm}\otimes \mathbf{1})\rho \otimes
\sigma (F_{nm}\otimes \mathbf{1})
\]
of the probabilities that Alice's measurement according to the observable $F$
will show the value $z_{nm}$.
\end{description}

The teleportation scheme works perfectly with respect to a certain class $%
\frak{S}$ of states $\rho $ on $\mathcal{H}_{1}$ if the following conditions
are fulfilled.

\begin{description}
\item[(E1)]  For each $n,m$ there exists a unitary operator $v_{nm}:\mathcal{%
H}_{1}\to \mathcal{H}_{3}$ such that
\[
\Lambda _{nm}(\rho )=v_{nm}\;\rho \;v_{nm}^{*}\quad (\rho \in \frak{S})
\]

\item[(E2)]
\[
\sum\limits_{nm}p_{nm}(\rho )=1\quad (\rho \in \frak{S})
\]

\item  (E1) means that Bob can reconstruct the original state $\rho $ by
unitary keys $\{v_{nm}\}$ provided to him. \newline

\item  (E2) means that Bob will succeed to find a proper key with certainty.
\newline
\end{description}

In the papers \cite{Ben,Ben2}, the authors used EPR spin pair to construct a
teleportation model. In order to have a more handy model, we here use
coherent states to construct a model. One of the main points for such a
construction is how to prepare the entangled state. The EPR entangled state
used in \cite{Ben} can be identified with the splitting of a one particle
state, so that the teleportation model of Bennett et al. can be described in
terms of Fock spaces and splittings, which makes us possible to work the
whole teleportation process in general beam splitting scheme. Moreover to
work with beams having a fixed number of particles seems to be not
realistic, especially in the case of large distance between Alice and Bob,
because we have to take into account that the beams will lose particles (or
energy). For that reason one should use a class of beams being insensitive
to this loss of particles. That and other arguments lead to superpositions
of coherent beams. \newline

In section 2 of this paper, we construct a teleportation model being perfect
in the sense of conditions (E1) and (E2), where we take the Boson Fock space
$\Gamma (L^{2}(G)):=\mathcal{H}_{1}=\mathcal{H}_{2}=\mathcal{H}_{3}$ with a
certain class $\rho $ of states on this Fock space.

In section 3 we consider a teleportation model where the entangled state $%
\sigma$ is given by the splitting of a superposition of certain coherent
states. Unfortunately this model doesn't work perfectly, that is, neither
(E2) nor (E1) hold. However this model is more realistic than that in the
section 2, and we show that this model provides a nice approximation to be
perfect. To estimate the difference between the perfect teleportation and
non-perfect teleportation, we add a further step in the teleportation scheme:

\begin{description}
\item[Step 5:]  Bob will perform a measurement on his part of the system
according to the projection
\[
F_{+}:=\mathbf{1}-|\mathrm{exp}(0)><\mathrm{exp}(0)|
\]
where $|\mathrm{exp}(0)><\mathrm{exp}(0)|$ denotes the vacuum state (the
coherent state with density $0$).
\end{description}

Then our new teleportation channels (we denote it again by $\Lambda _{nm}$)
have the form
\[
\Lambda _{nm}(\rho ):=\mathrm{tr}_{12}\frac{(F_{nm}\otimes F_{+})\rho
\otimes \sigma (F_{nm}\otimes F_{+})}{\mathrm{tr}_{123}(F_{nm}\otimes
F_{+})\rho \otimes \sigma (F_{nm}\otimes F_{+})}
\]
and the corresponding probabilities are
\[
p_{nm}(\rho ):=\mathrm{tr}_{123}(F_{nm}\otimes F_{+})\,\rho \otimes \sigma
(F_{nm}\otimes F_{+})
\]
For this teleportation scheme, (E1) is fulfilled. Furthermore we get
\[
\sum\limits_{nm}p_{nm}(\rho )=\frac{(1-e^{-\frac{d}{2}})^{2}}{1+(N-1)e^{-d}}
\quad \left( \rightarrow 1\text{ }(d\to +\infty )\right)
\]
Here $N$ denotes the dimension of the Hilbert space and $d$ is the
expectation value of the total number of particles (or energy) of the beam,
so that in the case of high density (or energy) $``d\to +\infty "$ of the
beam the model works perfectly.

Specializing this model we consider in section 4 the teleportation of all
states on a finite dimensional Hilbert space (through the space $\mathbf{R}%
^{k}$). Further specialization leads to a teleportation model where Alice
and Bob are spatially separated, that is, we have to teleport the
information given by the state of our finite dimensional Hilbert space from
one region $X_{1}\subseteq \mathbf{R}^{k}$ into another region $%
X_{2}\subseteq \mathbf{R}^{k}$ with $X_{1}\cap X_{2}=\emptyset $, and Alice
can only perform local measurements (inside of region $X_{1}$) as well as
Bob (inside of $X_{2}$). \newline

\section{Basic Notions and Notations}

First we collect some basic facts concerning the (symmetric) Fock space. We
will introduce the Fock space in a way adapted to the language of counting
measures. For details we refer to \cite{FF1,FF2,FFL,AO2,L} and other papers
cited in \cite{FFL}. \newline

Let $G$ be an arbitrary complete separable metric space. Further, let $\mu $
be a locally finite diffuse measure on $G$, i.e. $\mu (B)<+\infty $ for
bounded measurable subsets of $G$ and $\mu (\{x\})=0$ for all singletons $%
x\in G$. In order to describe the teleportation of states on a finite
dimensional Hilbert space through the $k$--dimensional space $\mathbf{R}^{k}$%
, especially we are concerned with the case
\begin{eqnarray*}
G &=&\mathbf{R}^{k}\times \{1,\ldots ,N\} \\
\mu &=&l\times \#
\end{eqnarray*}
where $l$ is the $k$--dimensional Lebesgue measure and $\#$ denotes the
counting measure on $\{1,\ldots ,N\}$. \newline

Now by $M=M(G)$ we denote the set of all finite counting measures on $G$.
Since $\varphi \in M$ can be written in the form $\varphi=\sum%
\limits_{j=1}^{n}\delta _{x_{j}}$ for some $n=0,1,2,\ldots $ and $x_{j}\in G$
(where $\delta _{x}$ denotes the Dirac measures corresponding to $x\in G$)
the elements of $M$ can be interpreted as finite (symmetric) point
configurations in $G$. We equip $M$ with its canonical $\sigma $--algebra $%
\frak{W}$ (cf. \cite{FF1}, \cite{FF2}) and we consider the measure $F$ by
setting
\[
F(Y):=\mathcal{X}_{Y}(O)+\sum\limits_{n\ge 1}\frac{1}{n!}\int\limits_{G^{n}}
\mathcal{X}_{Y}\left( \sum\limits_{j=1}^{n}\delta _{x_{j}}\right)
\mu^{n}(d[x_{1},\ldots ,x_{n}])(Y\in \frak{W})
\]
Hereby, $\mathcal{X}_{Y}$ denotes the indicator function of a set $Y$ and $O$
represents the empty configuration, i.~e., $O(G)=0$. Observe that $F$ is a $%
\sigma $--finite measure. \newline

Since $\mu $ was assumed to be diffuse one easily checks that $F$ is
concentrated on the set of a simple configurations (i.e., without multiple
points)
\[
\hat{M}:=\{\varphi \in M|\varphi (\{x\})\le 1\text{ for all }x\in G\}
\]

\begin{fdef}
\label{def1} $\mathcal{M}=\mathcal{M}(G):=L^{2}(M,\frak{W},F)$ is called the
(symmetric) Fock space over $G$.
\end{fdef}

In \cite{FF1} it was proved that $\mathcal{M}$ and the Boson Fock space $%
\Gamma (L^{2}(G))$ in the usual definition are isomorphic. \newline
For each $\Phi \in \mathcal{M}$ with $\Phi \neq 0$ we denote by $|\Phi >$
the corresponding normalized vector
\[
|\Phi >:=\frac{\Phi }{||\Phi ||}
\]
Further, $|\Phi ><\Phi |$ denotes the corresponding one--dimensional
projection, describing the pure state given by the normalized vector $|\Phi>$%
. Now, for each $n\ge 1$ let $\mathcal{M}^{\otimes n}$ be the $n$--fold
tensor product of the Hilbert space $\mathcal{M}$. Obviously, $\mathcal{M}%
^{\otimes n}$ can be identified with $L^{2}(M^{n},F^{n})$.

\begin{fdef}
\label{def2} For a given function $g:G\to \Bbb{C}$ the function
\end{fdef}

\noindent $\mathrm{exp}\;(g):M\to \Bbb{C}$ defined by
\[
\mathrm{exp}\;(g)\,(\varphi ):=\left\{
\begin{array}{lll}
1 & \text{ if } & \varphi =0 \\
\prod_{x\in G,\varphi \left( \left\{ x\right\} \right) >0}g(x) &  & otherwise
\end{array}
\right.
\]
is called exponential vector generated by $g$.

Observe that $\mathrm{exp}\;(g)\in \mathcal{M}$ if and only if $g\in L^{2}(G)
$ and one has in this case \newline
$||\mathrm{exp}\;(g)||^{2}=e^{\Vert g\Vert ^{2}}$ and $|\mathrm{exp}%
\;(g)>=e^{-\frac{1}{2}\Vert g\Vert ^{2}}\mathrm{exp}\;(g)$. The projection
\noindent \noindent $|\mathrm{exp}\;(g)><\mathrm{exp}\;(g)|$ is called the
coherent state corresponding to $g\in L^{2}(G)$. In the special case $%
g\equiv 0$ we get the vacuum state
\[
|\mathrm{exp}(0)>=\mathcal{X}_{\{0\}}\;.
\]
The linear span of the exponential vectors of $\mathcal{M}$ is dense in $%
\mathcal{M}$, so that bounded operators and certain unbounded operators can
be characterized by their actions on exponential vectors.

\begin{fdef}
\label{def3} The operator $D:\mathrm{dom}(D)\to \mathcal{M}^{\otimes 2}$
given on a dense domain $\mathrm{dom}(D)\subset \mathcal{M}$ containing the
exponential vectors from $\mathcal{M}$ by
\[
D\psi (\varphi _{1},\varphi _{2}):=\psi (\varphi _{1}+\varphi _{2})\quad
(\psi \in \mathrm{dom}(D),\,\varphi _{1},\varphi _{2}\in M)
\]
is called compound Malliavin derivative.
\end{fdef}

On exponential vectors $\mathrm{exp}\;(g)$ with $g\in L^{2}(G),$ one gets
immediately
\begin{equation}
D\;\mathrm{exp}\;(g)=\mathrm{exp}\;(g)\otimes \;\mathrm{exp}\;(g)  \label{1}
\end{equation}

\begin{fdef}
\label{def4} The operator $S:\mathrm{dom}(S)\to \mathcal{M}$ given on a
dense domain $\mathrm{dom}\;(S)\subset \mathcal{M}^{\otimes 2}$ containing
tensor products of exponential vectors by
\[
S\Phi (\varphi ):=\sum\limits_{\tilde{\varphi}\le \varphi }\Phi (\tilde{%
\varphi},\varphi -\tilde{\varphi})\quad (\Phi \in \mathrm{dom}(S),\;\varphi
\in M)
\]
is called compound Skorohod integral.
\end{fdef}

One gets
\begin{equation}
\langle D\psi ,\Phi \rangle _{\mathcal{M}^{\otimes 2}}=\langle \psi ,S\Phi
\rangle _{\mathcal{M}}\quad (\psi \in \mathrm{dom}(D),\; \Phi \in \mathrm{dom%
}(S))  \label{2}
\end{equation}
\begin{equation}
S(\mathrm{exp}\;(g)\otimes \mathrm{exp}\;(h))=\mathrm{exp}\;(g+h)\quad
(g,h\in L^{2}(G))  \label{3}
\end{equation}
For more details we refer to \cite{FW}.

\begin{fdef}
\label{def5} Let $T$ be a linear operator on $L^{2}(G)$ with $\Vert T\Vert
\le 1$. Then the operator $\Gamma (T)$ called second quantization of $T$ is
the (uniquely determined) bounded operator on $\mathcal{M}$ fulfilling
\[
\Gamma (T)\mathrm{exp}\;(g)=\mathrm{exp}\;(Tg)\quad (g\in L^{2}(G))
\]
\end{fdef}

Clearly, it holds
\begin{eqnarray}
\Gamma (T_{1})\Gamma (T_{2}) &=&\Gamma (T_{1}T_{2})  \label{4} \\
\Gamma (T^{*}) &=&\Gamma (T^{*})  \nonumber
\end{eqnarray}
It follows that $\Gamma (T)$ is an unitary operator on $\mathcal{M}$ if $T$
is an unitary operator on $L^{2}(G)$.

\begin{flemma}
\label{def6} Let $K_{1},K_{2}$ be linear operators on $L^{2}(G)$ with
property
\begin{equation}
K_{1}^{*}K_{1}+K_{2}^{*}K_{2}=\mathbf{1}\;.  \label{5}
\end{equation}
Then there exists exactly one isometry $\nu _{K_{1},K_{2}}$ from $\mathcal{M}
$ to $\mathcal{M}^{\otimes 2}=\mathcal{M}\otimes \mathcal{M}$ with
\begin{equation}
\nu _{K_{1},K_{2}}\mathrm{exp}\;(g)=\mathrm{exp}(K_{1}g)\otimes \mathrm{exp}%
(K_{2}g)\quad (g\in L^{2}(G))  \label{6}
\end{equation}
Further it holds
\begin{equation}
\nu _{K_{1},K_{2}}=(\Gamma (K_{1})\otimes \Gamma (K_{2}))D  \label{7}
\end{equation}
(at least on $\mathrm{dom}(D)$ but one has the unique extension). \newline
The adjoint $\nu _{K_{1},K_{2}}^{*}$ of $\nu _{K_{1},K_{2}}$ is
characterized by
\begin{equation}
\nu _{K_{1},K_{2}}^{*}(\mathrm{exp}\;(h)\otimes \mathrm{exp}\;(g))=\mathrm{%
exp}(K_{1}^{*}h+K_{2}^{*}g)\quad (g,h\in L^{2}(G))  \label{8}
\end{equation}
and it holds
\begin{equation}
\nu _{K_{1},K_{2}}^{*}=S(\Gamma (K_{1}^{*})\otimes \Gamma (K_{2}^{*}))
\label{9}
\end{equation}
\end{flemma}

\begin{frem}
From $K_{1},K_{2}$ we get a transition expectation $\xi
_{K_{1}K_{2}}:\mathcal{M}\otimes \mathcal{M}\to \mathcal{M}$, using $\nu
_{K_{1},K_{2}}$ and the lifting $\xi _{K_{1}K_{2}}^{*}$ may be interpreted
as a certain splitting (cf. \cite{AO2}).
\end{frem}

\noindent \textbf{Proof of \ref{def6}.} We consider the operator
\[
B:=S(\Gamma (K_{1}^{*})\otimes \Gamma (K_{2}^{*}))(\Gamma (K_{1})\otimes
\Gamma (K_{2}))D
\]
on the dense domain $\mathrm{dom}(B)\subseteq \mathcal{M}$ spanned by the
exponential vectors. Using 
(\ref{1}), (\ref{3}), (\ref{4}) and (\ref{5}) we get
\[
B\;\mathrm{exp}\;(g)=\mathrm{exp}\;(g)\quad (g\in L^{2}(G))\;.
\]
It follows that the bounded linear unique extension of $B$ onto $\mathcal{M}$
coincides with the unity on $\mathcal{M}$
\begin{equation}
B=\mathbf{1}\;.  \label{10}
\end{equation}
On the other hand, by equation (\ref{7}) at least on $\mathrm{dom}\;(D),$ an
operator $\nu _{K_{1},K_{2}}$ is defined. Using (\ref{2}) and (\ref{4}) we
obtain
\begin{eqnarray*}
\Vert \nu _{K_{1},K_{2}}\psi \Vert ^{2} &=&\langle \nu _{K_{1},K_{2}}\psi
,\nu _{K_{1},K_{2}}\psi \rangle \quad (\psi \in \mathrm{dom}\;(D)) \\
&=&\langle \psi ,B\psi \rangle ,
\end{eqnarray*}
which implies
\[
\Vert \nu _{K_{1},K_{2}}\psi \Vert ^{2}=\Vert \psi \Vert ^{2}\quad (\psi \in
\mathrm{dom}\;(D)).\;
\]
because of (\ref{10}). It follows that $\nu _{K_{1},K_{2}}$ can be uniquely
extended to a bounded operator on $\mathcal{M}$ with
\[
\Vert \nu _{K_{1},K_{2}}\psi \Vert =\Vert \psi \Vert \quad (\psi \in
\mathcal{M}).
\]
Now from (\ref{7}) we obtain (\ref{6}) using (\ref{1}) and the definition of
the operators of second quantization. Further, (\ref{7}), (\ref{3}) and (\ref
{4}) imply (\ref{9}) and from (\ref{9}) we obtain (\ref{8}) using the
definition of the operators of second quantization and equation (\ref{3}). $%
\blacksquare $

Here we explain fundamental scheme of beam splitting \cite{FFL}. We define
an isometric operator $V_{\alpha ,\beta }$ for coherent vectors such that
\[
V_{\alpha ,\beta }|\,\mathrm{exp}\;(g)\rangle =|\,\mathrm{exp}\;(\alpha
g)\rangle \otimes |\,\mathrm{exp}\,(\beta g)\rangle
\]
with $\mid \alpha \mid ^{2}+\mid \beta \mid ^{2}=1$. This beam splitting is
a useful mathematical expression for optical communication and quantum
measurements \cite{AO2}.

\begin{fex}
\label{def7} $\left( \alpha =\beta =1/\sqrt{2}\text{ above}\right) $ Let $%
K_{1}=K_{2}$ be the following operator of multiplication on $L^{2}(G)$
\[
K_{1}g=\frac{1}{\sqrt{2}}\;g=K_{2}g\quad (g\in L^{2}(G))
\]
We put
\[
\nu :=\nu _{K_{1},K_{2}}
\]
and obtain
\[
\nu \;\mathrm{exp}\;(g)=\mathrm{exp}\;\left( \frac{1}{\sqrt{2}}g\right)
\otimes \mathrm{exp}\;(\frac{1}{\sqrt{2}}\;g)\quad (g\in L^{2}(G))
\]
\end{fex}

\begin{fex}
\label{def8} Let $L^{2}(G)=\mathcal{H}_{1}\oplus \mathcal{H}_{2}$ be the
orthogonal sum of the subspaces $\mathcal{H}_{1},\mathcal{H}_{2}$. $K_{1}$
and $K_{2}$ denote the corresponding projections.
\end{fex}

We will use Example \ref{def7} in order to describe a teleportation model
where Bob performs his experiments on the same ensemble of the systems like
Alice. \newline

Further we will use a special case of Example \ref{def8} in order to
describe a teleportation model where Bob and Alice are spatially separated
(cf. section 5).

\begin{frem}
\label{def9} The property (\ref{5}) implies
\begin{equation}
\Vert K_{1}g\Vert ^{2}+\Vert K_{2}g\Vert ^{2}=\Vert g\Vert ^{2}\quad (g\in
L^{2}(G))  \label{11}
\end{equation}
\end{frem}

\begin{frem}
\label{def10} Let $U$, $V$ be unitary operators on $L^{2}(G)$. If operators $%
K_{1},K_{2}$ satisfy (\ref{5}),~then the pair $\hat{K}_{1}=UK_{1},\ \hat{K}%
_{2}=VK_{2}$ fulfill (\ref{5}).
\end{frem}

\section{A perfect model of teleportation}

\label{sec2} Concerning the general idea we follow the papers \cite{IOS},
\cite{AO}. We fix an ONS $\{g_{1},\ldots ,g_{N}\}\subseteq L^{2}(G)$,
operators $K_{1},K_{2}$ on $L^{2}(G)$ with (\ref{5}), an unitary operator $T$
on $L^{2}(G)$, and $d>0$. We assume
\begin{equation}
TK_{1}g_{k}=K_{2}g_{k}\quad (k=1,\ldots ,N),  \label{12}
\end{equation}
\begin{equation}
\langle K_{1}g_{k},K_{1}g_{j}\rangle =0\quad (k\not{=}j;\;k,j=1\ldots ,N),
\label{13}
\end{equation}
Using (\ref{11}) and (\ref{12}) we get
\begin{equation}
\Vert K_{1}g_{k}\Vert ^{2}=\Vert K_{2}g_{k}\Vert ^{2}=\frac{1}{2}.
\label{14}
\end{equation}
From (\ref{12}) and (\ref{13}) we get
\begin{equation}
\langle K_{2}g_{k},\,K_{2}g_{j}\rangle =0\quad (k\neq j\,;\;k,j=1,\ldots ,N).
\label{15}
\end{equation}
The state of Alice asked to teleport is of the type
\begin{equation}
\rho =\sum\limits_{s=1}^{N}\lambda _{s}|\Phi _{s}\rangle \langle \Phi _{s}|,
\label{16}
\end{equation}
where
\begin{equation}
|\Phi _{s}\rangle =\sum\limits_{j=1}^{N}c_{sj}|\mathrm{exp}\;(aK_{1}g_{j})-%
\mathrm{exp}\;(0)\rangle \quad
\left(\sum\limits_{j}|c_{sj}|^{2}=1;s=1,\ldots ,N\right)  \label{17}
\end{equation}
and $a=\sqrt{d}$. One easily checks that $(|\mathrm{exp}\;(aK_{1}g_{j})-%
\mathrm{exp}\;(0)\rangle )_{j=1}^{N}$ and $(|\mathrm{exp}\;aK_{2}g_{j})-%
\mathrm{exp}\;(0)\rangle )_{j=1}^{N}$ are ONS in $\mathcal{M}$. \newline

In order to achieve that $(|\Phi _{s}\rangle )_{s=1}^{N}$ is still an ONS in
$\mathcal{M}$ we assume
\begin{equation}
\sum\limits_{j=1}^{N}\bar{c}_{sj}c_{kj}=0\quad (j\neq
k\,;\;j,k=1,\ldots,N)\,.  \label{18}
\end{equation}
Denote $c_{s}=[c_{s1,\ldots ,}c_{sN}]\in \Bbb{C}^{N}$, then $%
(c_{s})_{s=1}^{N}$ is an CONS in $\Bbb{C}^{N}$.

Now let $(b_{n})_{n=1}^{N}$ be a sequence in $\Bbb{C}^{N}$,
\[
b_{n}=[b_{n1,\ldots ,}b_{nN}]
\]
with properties
\begin{equation}
|b_{nk}|=1\quad (n,k=1,\ldots ,N),  \label{19}
\end{equation}
\begin{equation}
\langle b_{n}\,,\;b_{j}\rangle =0\quad (n\neq j\,;\;n,j=1,\ldots ,N).
\label{20}
\end{equation}
Then Alice's measurements are performed with projection
\begin{equation}
F_{nm}=|\xi _{nm}\rangle \langle \xi _{nm}|\quad (n,m=1,\ldots ,N)
\label{21}
\end{equation}
given by
\begin{equation}
|\xi _{nm}\rangle =\frac{1}{\sqrt{N}}\sum\limits_{j=1}^{N}b_{nj}|\mathrm{exp}%
\;(aK_{1}g_{j})-\mathrm{exp}\;(0)>\otimes |\;\mathrm{exp}\;(aK_{1}g_{j\oplus
m})-\mathrm{exp}\;(0)\rangle ,  \label{22}
\end{equation}
where $j\oplus m:=j+m(\mathrm{mod}\;N)$. \newline

One easily checks that $(|\xi _{nm}\rangle )_{n,m=1}^{N}$ is an ONS in $%
\mathcal{M}^{\otimes 2}$. Further, the state vector $|\xi \rangle $ of the
entangled state $\sigma =|\xi \rangle \langle \xi |$ is given by
\begin{equation}
|\xi \rangle =\frac{1}{\sqrt{N}}\sum\limits_{k}|\mathrm{exp}\;(aK_{1}g_{k})-
\mathrm{exp}\;(0)\rangle \otimes |\mathrm{exp}\;(aK_{2}g_{k})-\mathrm{exp}\;
(0)\rangle \,.  \label{23}
\end{equation}

\begin{flemma}
\label{def11} For each $n,m=1,\ldots ,N$ it holds
\begin{eqnarray}
&&(F_{nm}\otimes \mathbf{1})(|\Phi _{s}\rangle \otimes |\xi \rangle )
\nonumber \\
&=&\frac{1}{N}|\xi _{nm}\rangle \otimes \sum\limits_{j}\bar{b}_{nj}c_{sj}|%
\mathrm{exp}\;(aK_{2}g_{j\oplus m})-\mathrm{exp}\;(0)\rangle \text{ }%
(s=1,\ldots ,N)  \label{24}
\end{eqnarray}
\end{flemma}

\noindent \textbf{Proof: }From the fact that
\begin{equation}
|\gamma _{j}\rangle :=|\mathrm{exp}\;(aK_{1}g_{j})-\mathrm{exp}\;(0)\rangle
\quad \left( j=1,\ldots ,N\right)  \label{25}
\end{equation}
is an ONS, it follows
\begin{equation}
\langle \gamma _{\emph{r}}\otimes \gamma _{\emph{r}\oplus m}\,,\;
\gamma_{j}\otimes \gamma _{k}\rangle = \left\{
\begin{array}{ll}
1 & \text{if }\emph{r}=j\text{ and }k=\emph{r}\oplus m \\
0 & \text{otherwise}
\end{array}
\right. .  \label{26}
\end{equation}
On the other hand, we have
\begin{eqnarray}
&&(F_{nm}\otimes \mathbf{1})(|\Phi _{s}\rangle \otimes |\xi \rangle )
\nonumber \\
&=&\frac{1}{N}\sum\limits_{k}\sum\limits_{j}\sum\limits_{\emph{r}}c_{sj}\bar
{b}_{ns}\langle \gamma _{\emph{r}}\otimes \gamma _{\emph{r}\oplus m}\,,\;
\gamma _{j}\otimes \gamma _{k}\rangle \xi _{nm}\otimes |\mathrm{exp}\;
(aK_{2}g_{k})-\mathrm{exp}\;(0)\rangle .  \nonumber \\
&&  \label{27}
\end{eqnarray}
\noindent Using (\ref{26}) and (\ref{27}), we get (\ref{24}) \hfill $%
\blacksquare$ \newline

Now we have

\begin{eqnarray}
\rho \otimes \sigma &=&\sum\limits_{s=1}^{N}\lambda _{s}|\Phi _{s}\rangle
\langle \Phi _{s}|\otimes |\xi \rangle \langle \xi |  \label{28} \\
&=&\sum\limits_{s=1}^{N}\lambda _{s}|\Phi _{s}\otimes \xi \rangle \langle
\Phi _{s}\otimes \xi |\;,  \nonumber
\end{eqnarray}
which implies
\begin{eqnarray}
(F_{nm}\otimes \mathbf{1})(\rho \otimes \sigma )(F_{nm}\otimes \mathbf{1})
&=&\sum\limits_{s=1}^{N}\lambda _{s}(F_{nm}\otimes \mathbf{1}%
)|\Phi_{s}\otimes \xi \rangle \langle \Phi _{s}\otimes \xi |(F_{nm}\otimes
\mathbf{\ 1})  \nonumber \\
&=&\sum\limits_{s=1}^{N}\lambda _{s}\Vert (F_{nm}\otimes \mathbf{1})
(\Phi_{s}\otimes \xi )\Vert ^{2}  \label{29} \\
&&|(F_{nm}\otimes \mathbf{1})(\Phi _{s}\otimes \xi )\rangle \langle
(F_{nm}\otimes \mathbf{1})(\Phi _{s}\otimes \xi )|.  \nonumber
\end{eqnarray}
Note $|\Phi _{s}\otimes \xi \rangle =|\Phi _{s}\rangle \otimes |\xi \rangle $%
. From (\ref{12}) it follows that
\begin{equation}
\sum\limits_{j}\bar{b}_{nj}c_{sj}|\mathrm{exp}\;(aK_{2}g_{j\oplus m})-
\mathrm{exp}\;(0)\rangle =\Gamma (T)\sum\limits_{j}\bar{b}_{nj}c_{sj}|%
\mathrm{exp} \;(aK_{1}g_{j\oplus m})-\mathrm{exp}\;(0)\rangle .  \label{30}
\end{equation}
Further, for each $m,n\left( =1,\ldots ,N\right) ,$ we have unitary
operators $U_{m},B_{n}$ on $\mathcal{M}$ given by
\begin{equation}
B_{n}|\mathrm{exp}\;(aK_{1}g_{j})-\mathrm{exp}\;(0)\rangle =b_{nj}|\mathrm{e
xp}\;(aK_{1}g_{j})-\mathrm{exp}\;(0)\rangle \quad (j=1,\ldots ,N)  \label{31}
\end{equation}
\begin{equation}
U_{m}|\mathrm{exp}\;(aK_{1}g_{j})-\mathrm{exp}\;(0)\rangle =|\mathrm{exp}\;
(aK_{1}g_{j\oplus m})-\mathrm{exp}\;(0)\rangle \quad (j=1,\ldots ,N)
\label{32}
\end{equation}
Therefore we get
\begin{equation}
\sum\limits_{j}\bar{b}_{nj}c_{sj}|\mathrm{exp}\;(aK_{1}g_{j\oplus m})-%
\mathrm{exp}\;(0)\rangle =U_{m}B_{n}^{*}(\Phi _{s})  \label{33}
\end{equation}
From (\ref{30}), (\ref{33}) and Lemma \ref{def11} we obtain
\begin{equation}
\left( F_{nm}\otimes \mathbf{1}\right) \left( |\Phi _{s}\rangle \otimes |\xi
\rangle \right) =\frac{1}{N}|\xi _{nm}\rangle \otimes \left( \Gamma
(T)U_{m}B_{n}^{*}|\Phi _{s}\rangle \right)  \label{34}
\end{equation}
It follows
\begin{equation}
\Vert \left( F_{nm}\otimes \mathbf{1}\right) \left( |\Phi _{s}\rangle
\otimes |\xi \rangle \right) \Vert ^{2}=\frac{1}{N^{2}}  \label{35}
\end{equation}
Finally from (\ref{29}), (\ref{34}) and (\ref{35}) we have
\begin{equation}
\left( F_{nm}\otimes \mathbf{1}\right) (\rho \otimes \sigma )
\left(F_{nm}\otimes \mathbf{1}\right) =\frac{1}{N^{2}}F_{nm}\otimes \left(
\Gamma(T)U_{m}B_{n}^{*}\right) \rho \left( B_{n}U_{m}^{*}\Gamma
(T^{*})\right)  \label{36}
\end{equation}
That leads to the following solution of the teleportation problem.

\begin{ftheo}
\label{def12} For each $n,m=1,\ldots ,N$, define a channel $\Lambda _{nm}$
by
\begin{equation}
\Lambda _{nm}(\rho ):=\mathrm{tr}_{12}\frac{\left( F_{nm}\otimes 1\right)
(\rho \otimes \sigma )\left( F_{nm}\otimes \mathbf{1}\right) }{\mathrm{tr}%
_{123}\left( F_{nm}\otimes 1\right) \left( \hat{\rho}\otimes \sigma \right)
\left( F_{nm}\otimes \mathbf{1}\right) }\quad (\rho \text{ normal state on }%
\mathcal{M})  \label{37}
\end{equation}
Then we have for all states $\rho $ on $M$ with (\ref{16}) and (\ref{17})
\begin{equation}
\Lambda _{nm}(\rho )=\left( \Gamma (T)U_{m}B_{n}^{*}\right) \rho \left(
\Gamma (T)U_{m}B_{n}^{*}\right) ^{*}  \label{38}
\end{equation}
\end{ftheo}

\begin{frem}
\label{def13} In case of Example \ref{def7} using the operators $%
B_{n},U_{m},\Gamma (T),$ the projections $F_{nm}$ are given by unitary
transformations of the entangled state $\sigma $:
\begin{eqnarray}
F_{nm} &=&\left( B_{n}\otimes U_{m}\Gamma (T^{*})\right) \sigma \left(
B_{n}\otimes U_{m}\Gamma (T^{*})\right) ^{*}  \label{39} \\
&&\text{or}\hspace*{2cm}  \nonumber \\
|\xi _{nm}\rangle  &=&\left( B_{n}\otimes U_{m}\Gamma (T^{*})\right) |\xi
\rangle   \nonumber
\end{eqnarray}
\end{frem}

\begin{frem}
\label{def14} If Alice performs a measurement according to the following
selfadjoint operator
\[
F=\sum\limits_{n,m=1}^{N}z_{nm}F_{nm}
\]
with $\{z_{nm}|n,m=1,\ldots ,N\}\subseteq \mathbf{R}-\{0\},$ then she will
obtain the value $z_{nm}$ with probability $1/N^{2}$. The sum over all this
probabilities is $1$, so that the teleportation model works perfectly.
\end{frem}

\section{A non--perfect case of Teleportation}

In this section we will construct a model where we have also channels with
property (\ref{38}). But the probability that one of these channels will
work in order to teleport the state from Alice to Bob is less than $1$
depending on the density parameter $d$ (or energy of the beams, depending on
the interpretation). If $d=a^{2}$ tends to infinity that probability tends
to $1$. That is the model is asymptotically perfect in a certain sense.

We consider the normalized vector
\begin{eqnarray}
|\eta \rangle := &&\frac{\gamma }{\sqrt{N}}\sum\limits_{k=1}^{N}|\mathrm{exp
}\;(ag_{k})\rangle  \label{40} \\
\gamma := &&\left( \frac{1}{1+(N-1)e^{-d}}\right) ^{\frac{1}{2}}=\left(
\frac{1}{1+(N-1)e^{-a^{2}}}\right) ^{\frac{1}{2}}  \nonumber
\end{eqnarray}
and we replace in (\ref{37}) the projector $\sigma $ by the projector
\begin{eqnarray}
\tilde{\sigma}:= &&|\tilde{\xi}\rangle \langle \tilde{\xi}|  \label{41} \\
\tilde{\xi}:= &&\nu _{K_{1},K_{2}}(\eta )=\frac{\gamma }{\sqrt{N}}
\sum\limits_{k=1}^{N}|\mathrm{exp}\;(aK_{1}g_{k})\rangle \otimes |\mathrm{ex
p}\;(aK_{2}g_{k})\rangle  \nonumber
\end{eqnarray}
Then for each $n,m=1,\ldots ,N,$ we get the channels on a normal state $\rho
$ on $\mathcal{M}$ such as
\begin{eqnarray}
\tilde{\Lambda}_{nm}(\rho ):= &&\mathrm{tr}_{12}\frac{\left( F_{nm}\otimes
\mathbf{1}\right) \left( \rho \otimes \tilde{\sigma}\right) \left(
F_{nm}\otimes \mathbf{1}\right) }{\mathrm{tr}_{123}\left( F_{nm}\otimes
\mathbf{1}\right) \left( \rho \otimes \tilde{\sigma}\right) \left(
F_{nm}\otimes \mathbf{1}\right) }\quad  \label{42} \\[0.12in]
\Theta _{nm}(\rho ):= &&\mathrm{tr}_{12}\frac{\left( F_{nm}\otimes
F_{+}\right) \left( \rho \otimes \tilde{\sigma}\right) \left( F_{nm}\otimes
F_{+}\right) }{\mathrm{tr}_{123}\left( F_{nm}\otimes F_{+}\right) \left(
\rho \otimes \tilde{\sigma}\right) \left( F_{nm}\otimes F_{+}\right) }\;,
\label{43}
\end{eqnarray}
where $F_{+}=\mathbf{1}-|\mathrm{exp}\;(0)\rangle \langle \mathrm{exp}\;(0)|$
e.~g., $F_{+}$ is the projection onto the space $\mathcal{M}_{+}$ of
configurations having no vacuum part, e.~g., orthogonal to vacuum
\[
\mathcal{M}_{+}:=\{\psi \in \mathcal{M}|\;\Vert \mathrm{exp}\;(0)\rangle
\langle \mathrm{exp}\;(0)|\psi \Vert =0\}
\]
One easily checks that
\begin{equation}
\Theta _{nm}(\rho )=\frac{F_{+}\tilde{\Lambda}_{nm}(\rho )F_{+}}{\mathrm{tr}
\left( F_{+}\tilde{\Lambda}_{nm}(\rho )F_{+}\right) }  \label{44}
\end{equation}
that is, after receiving the state $\tilde{\Lambda}_{nm}(\rho )$ from Alice,
Bob has to omit the vacuum. \newline

From Theorem \ref{def12} it follows that for all $\rho $ with (\ref{16})
and (\ref{17})
\[
\Lambda _{nm}(\rho )=\frac{F_{+}\Lambda _{nm}(\rho )F_{+}}{\mathrm{tr}\;
(F_{+}\Lambda _{nm}(\rho )F_{+})}
\]
This is not true if we replace $\Lambda _{nm}$ by $\tilde{\Lambda}_{nm}$,
namely, in general it does not hold
\[
\Theta _{nm}(\rho )=\tilde{\Lambda}_{nm}(\rho )
\]
But we will prove that for each $\rho $ with (\ref{16}), and (\ref{17}) it
holds

\[
\Theta _{nm}(\rho )=\Lambda_{nm}(\rho )
\]

\noindent which means
\begin{equation}
\Theta _{nm}(\rho )=(\Gamma (T)U_{m}B_{n}^{*})\rho (\Gamma
(T)U_{m}B_{n}^{*})^{*}  \label{45}
\end{equation}
because of Theorem \ref{def12}. Further we will show
\begin{equation}
\mathrm{tr_{123}}\left( F_{nm}\otimes F_{+}\right) \left( \rho \otimes
\tilde{\sigma}\right) \left( F_{nm}\otimes F_{+}\right) =\frac{\gamma ^{2}}{%
N^{2}}\left( e^{\frac{d}{2}}-1\right) ^{2}e^{-d}  \label{46}
\end{equation}
and the sum over $n,m\left( =1,\ldots ,N\right) $ gives the probability
\[
\frac{\left( 1-e^{-\frac{d}{2}}\right) ^{2}}{1+(N-1)e^{-d}}\longrightarrow 1%
\text{ }\left( d\longrightarrow \infty \right)
\]
which means that the teleportation model works perfectly in the limit $%
d\longrightarrow \infty $, e.~g., Bob will receive one of the states $\Theta
_{nm}(\rho )$ given by (\ref{44}). Thus we formulate the following theorem.

\begin{ftheo}
\label{def15} For all states $\rho $ on $\mathcal{M}$ with (\ref{16}) and (%
\ref{17}) and each pair $n,m\left( =1,\ldots ,N\right) ,$ the equations (\ref
{44}) and (\ref{45}) hold. Further, we have
\begin{equation}
\sum\limits_{n,m}\mathrm{tr}_{123}\left( F_{nm}\otimes F_{+}\right) \left(
\rho \otimes \tilde{\sigma}\right) \left( F_{nm}\otimes F_{+}\right) =\frac{%
\left( 1-e^{-\frac{d}{2}}\right) ^{2}}{1+(N-1)e^{-d}}.  \label{47}
\end{equation}
\end{ftheo}

In order to prove theorem \ref{def15}, we fix $\rho $ with (\ref{16}) and (%
\ref{17}) and start with a lemma.

\begin{flemma}
\label{def16} For each $n,m,s\left( =1,\ldots ,N\right) ,$ it holds
\begin{eqnarray*}
\left( F_{nm}\otimes \mathbf{1}\right) \left( |\Phi _{s}\rangle \otimes |%
\tilde{\xi}\rangle \right)  &=&\frac{\gamma }{N}\left( 1-e^{-\frac{d}{2}%
}\right) |\xi _{nm}\rangle \otimes \left( \Gamma (T)U_{m}B_{n}^{*}|\Phi
_{s}\rangle \right)  \\
&&+\frac{\gamma }{N}\left( \frac{e^{\frac{d}{2}}-1}{e^{d}}\right) ^{\frac{1}{%
2}}\langle b_{n},c_{s}\rangle _{\Bbb{C}^{N}}\xi _{nm}\otimes |\mathrm{exp}%
\;(0)\rangle
\end{eqnarray*}
\end{flemma}

\noindent \textbf{Proof:} For all $k,j,\emph{r}=1,\ldots ,N,$ we get
\begin{eqnarray*}
\alpha _{k,j,\emph{r}}:= && \langle |\mathrm{exp}\;(aK_{1}g_{\emph{r}})-
\mathrm{exp}\;(0)\rangle \otimes ||\mathrm{exp}\; (aK_{1}g_{\emph{r}\otimes
m})-\mathrm{exp}\;(0)\rangle \,, \\
&&~|\mathrm{exp}\;(aK_{1}g_{j})-\mathrm{exp}\;(0)\rangle \otimes |\mathrm{exp%
}\;(aK_{1}g_{k})\rangle \rangle \\
&=&\left\{
\begin{array}{ll}
\left( \frac{e^{\frac{a^{2}}{2}}-1}{e^{\frac{a^{2}}{2}}}\right) & \text{if }%
\emph{r}=j\text{ and }k=\emph{r}\oplus m \\
0 & \text{otherwise}
\end{array}
\right.
\end{eqnarray*}
and
\[
|\mathrm{exp}\left( aK_{2}g_{j\oplus m}\right) \rangle =e^{-\frac{a^{2}}{2}}
\left( e^{\frac{a^{2}}{2}}-1\right) ^{\frac{1}{2}}|\mathrm{exp}\;\left(
aK_{2}g_{j\oplus m}\right) -\mathrm{exp}\;(0)\rangle +e^{-\frac{a^{2}}{2}}|
\mathrm{exp}\;(0)\rangle
\]
On the other hand, we have
\[
(F_{nm}\otimes \mathbf{1})\left( |\Phi _{s}\rangle \otimes |\tilde{\xi}
\rangle \right) =\frac{\gamma }{N}\sum\limits_{k}\sum\limits_{j}\sum\limits_
{\emph{r}}c_{sj}\bar{b}_{nr}\alpha _{k,j,\emph{r}}\xi _{nm}\otimes |\mathrm{%
\ exp}\;(aK_{2}g_{k})\rangle
\]
It follows with $a^{2}=d$
\begin{eqnarray*}
\left( F_{nm}\otimes \mathbf{1}\right) \left( \Phi _{s}\otimes \tilde{\xi}
\right) &=& \frac{\gamma }{N}\left( e^{\frac{d}{2}}-1\right) e^{-\frac{d}{2}
}\xi _{nm}\otimes \left( \sum\limits_{j}c_{sj}\bar{b}_{nj}|\mathrm{exp}\;
\left( aK_{2}g_{j\oplus m}\right) -\mathrm{exp}\;(0)\rangle \right) \\
&&+\frac{\gamma }{N}\left( e^{\frac{d}{2}}-1\right) ^{\frac{1}{2}} e^{-\frac{%
d}{2}}\sum\limits_{j}c_{sj}\bar{b}_{nj}\xi _{nm} \otimes |\mathrm{exp}%
\;(0)\rangle \\
&=& \frac{\gamma }{N}\left( 1-e^{-\frac{d}{2}}\right) \xi _{nm}\otimes
\left(\Gamma (T)U_{m}B_{n}^{*}\Phi _{s}\right) \\
&& +\frac{\gamma }{N} \left( \frac{e^{\frac{d}{2}}-1}{e^{d}}\right) ^{\frac{1%
}{2}}\langle b_{n},c_{s} \rangle _{\Bbb{C}^{N}}\xi _{nm}\otimes |\mathrm{exp}%
\; (0)\rangle .\text{ }\blacksquare
\end{eqnarray*}
\hfill\newline
\label{def17} If $\rho $ is a pure state
\[
\rho =|\Phi _{s}\rangle \langle \Phi _{s}|
\]
then we obtain from Lemma \ref{def16}
\[
\begin{array}{ll}
& \mathrm{tr}_{123}\left( F_{nm}\otimes \mathbf{1}\right) \left( \rho
\otimes \tilde{\sigma}\right) \left( F_{nm}\otimes \mathbf{1}\right) \\
& =\frac{\gamma ^{2}}{N^{2}}\left( \left( 1-e^{-\frac{d}{2}}\right) ^{2}+
\frac{e^{\frac{d}{2}}-1}{e^{d}}|\langle b_{n},c_{s}\rangle |^{2}\right) \\
& =\frac{1}{N^{2}\left( 1+(N-1)e^{-d}\right) } \left( \left( 1-e^{-\frac{d}{
2%
}}\right) ^{2}+\frac{e^{\frac{d}{2}}-1}{e^{d}}|\langle
b_{n},c_{s}\rangle|^{2}\right)
\end{array}
\]
and
\[
\tilde{\Lambda}_{nm}(\rho )\not{=}\left( \Gamma (T)U_{m}B_{n}^{*}\right)
\rho \left( \Gamma (T)U_{m}B_{n}^{*}\right) ^{*}.
\]
\medskip Now we have \noindent \vspace{0cm}
\[
\Gamma (T)U_{m}B_{n}^{*}\Phi _{s}\in \mathcal{M}_{+}\,,\;|\mathrm{exp}\;(0)
\rangle \in \mathcal{M}_{+}^{\bot }
\]
Hence, Lemma \ref{def16} implies
\[
\left( \mathbf{1}\otimes \mathbf{1}\otimes F_{+}\right) \left( F_{nm}\otimes
\mathbf{1}\right) \left( \Phi _{s}\otimes \tilde{\xi}\right) =\frac{\gamma
} {N}\left( 1-e^{-\frac{d}{2}}\right) \xi _{nm}\otimes \left( \Gamma
(T)U_{m}B_{n}^{*}\Phi _{s}\right)
\]
that is, we have the following Lemma

\begin{flemma}
\label{def18} For each $n,m,s=1,\ldots ,N,$ it holds
\begin{equation}
\left( F_{nm}\otimes F_{+}\right) \left( \Phi _{s}\otimes \tilde{\xi}\right)
=\frac{\gamma }{N}\left( 1-e^{-\frac{d}{2}}\right) \xi _{nm}\otimes \left(
\Gamma (T)U_{m}B_{n}^{*}\Phi _{s}\right) .  \label{48}
\end{equation}
\end{flemma}

\begin{frem}
\label{def19} Let $K_{2}$ be a projection of the type
\[
K_{2}h=h\mathcal{X}_{X};\text{ }h\in L^{2}(G),
\]
where $X\subseteq G$ is measurable. Then (\ref{48}) also holds if we replace
$F_{+}$ by the projection $F_{+,X}$ onto the subspace $\mathcal{M}_{+,X}$ of
$\mathcal{M}$ given by
\[
\mathcal{M}_{+,X}:=\{\psi \in \mathcal{M}|\psi (\varphi )=0\text{ if }%
\varphi (X)=0\}
\]
Observe that $\mathcal{M}_{+,G}=\mathcal{M}_{+}$.
\end{frem}

\noindent \textbf{Proof of theorem \ref{def15}}: We have assumed that $%
\left
( |\Phi _{s}\rangle \right) _{s=1}^{N}$ is an ONS in $\mathcal{M}$,
which implies that $\left( |\xi _{nm}\rangle \otimes \left( \Gamma
(T)U_{m}B_{n}^{*}|\Phi _{s}\rangle \right) \right) _{s=1}^{N}$ is an ONS in $%
\mathcal{M}^{\otimes 3}$. Hence we obtain the equations (\ref{45}), (\ref{46}%
) and (\ref{47}) by Lemma \ref{def18}. This proves Theorem \ref{def15}. $%
\blacksquare $

\begin{frem}
\label{def20} In the special case of the remark \ref{def19}, the equations (
\ref{45}), (\ref{46}) and (\ref{47}) hold if we replace $F_{+}$ by $F_{+,X}$
in the definition of the channel $\Theta _{nm}$ and in (\ref{46}), (\ref{47}
), that is, Bob will only perform ``local'' measurement according to the
region $X$, about which we will discuss more details in the next sections.
\end{frem}

\section{Teleportation of states inside $\mathbf{R}^{k}$}

\label{sec4}Let $\mathcal{H}$ be a finite-dimensional Hilbert space. We
consider the case $\mathcal{H}=\Bbb{C}^{N}=L^{2}(\{1,\ldots ,N\},\#)$
without loss of generality, where $\#$ denotes the counting measure on the
set $\{1,\ldots ,N\}$. We want to teleport states on $\mathcal{H}$ with the
aid of the constructed channels $\left( \Lambda _{nm}\right) _{n,m=1}^{N}$
or $\left( \Theta _{nm}\right) _{n,m=1}^{N}$. We fix

\begin{enumerate}
\item[-]  a CONS $(|j\rangle )_{j=1}^{N}$ of $\mathcal{H}$

\item[-]  $f\in L^{2}\left( \mathbf{R}^{k}\right) $, $\Vert f\Vert =1$

\item[-]  $d=a^{2}>0$

\item[-]  $\hat{K}_{1},\hat{K}_{2}$ linear operators on $L^{2}\left( \mathbf{%
R}^{k}\right) $

\item[-]  $\hat{T}$ unitary operator on $L^{2}\left( \mathbf{R}^{k}\right) $
\end{enumerate}

with two properties
\begin{equation}
\hat{K}_{1}^{*}\hat{K}_{1}f+\hat{K}_{2}^{*}\hat{K}_{2}f=f  \label{49}
\end{equation}
\begin{equation}
\hat{T}\hat{K}_{1}f=\hat{K}_{2}f  \label{50}
\end{equation}
We put
\[
G=\mathbf{R}^{k}\times \{1,\ldots ,N\}\,,\;\mu =l\times \#,
\]
where $l$ is the Lebesgues measure on $\mathbf{R}^{k}$. Then $L^{2}(G)=L^{2}
(G,\mu )=L^{2}(\mathbf{R}^{k})\otimes \mathcal{H}.$ Further, put
\[
g_{j}:=f\otimes |j\rangle \quad (j=1,\ldots ,N)
\]
Then $(g_{j})_{j=1}^{N}$ is an ONS in $L^{2}(G)$. We consider linear
operators $K_{1},K_{2}$ on $L^{2}(G)$ with (\ref{5}) and
\begin{equation}
K_{\emph{r}}g_{j}=\left( \hat{K}_{\emph{r}}f\right) \otimes |j\rangle \quad
(j=1,\ldots ,N;\;\emph{r}=1,2).  \label{51}
\end{equation}

\begin{frem}
\label{def21} (\ref{51}) determines operators $K_{1},K_{2}$ on the subspace
of $\mathcal{M}$ spanned by the ONS $(g_{j})_{j=1}^{N}$. On the orthogonal
complement, one can put for instance
\[
K_{\emph{r}}\psi =\frac{1}{\sqrt{2}}\psi
\]
Then $K_{1},K_{2}$ are well defined and fulfill (\ref{5}) because of (\ref
{49}). Further, one checks that (\ref{13}) and (\ref{15}) hold.
\end{frem}

Now let $T$ be an unitary operator on $L^{2}(G)$ with
\[
T(K_{1}g_{j})=\left( \hat{T}\hat{K}_{2}f\right) \otimes |j\rangle
\]
From (\ref{13}) one can prove the existence of $T$ using the arguments as
in the remark \ref{def21}. Further, we get (\ref{12}) from (\ref{50}).
\newline
Summarizing, we obtain that $\{g_{1},\ldots ,g_{N}\}$, $K_{1},K_{2},T$
fulfill all the assumptions required in section 2. Thus we have the
corresponding channels $\Lambda _{n,m}$ ,$\Theta _{nm}$ given by (\ref{37})
and (\ref{43}) respectively. It follows that we are able to teleport a state
$\rho $ on $\mathcal{M}=\mathcal{M}(G)$ with (\ref{16}) and (\ref{17} ) as
it was stated in the theorem \ref{def12} and the theorem \ref{def15},
respectively. \newline
In order to teleport states on $\mathcal{H}$ through the space $\mathbf{R}%
^{k}$ using the above channels, we have to consider:

\begin{description}
\item[first:]  a ``lifting'' $\mathcal{E}^{*}$ of the states on $\mathcal{H}$
into the set of states on the bigger state space on $\mathcal{M}$ such that $%
\rho =\mathcal{E}^{*}(\hat{\rho})$ can be described by (\ref{16}), (\ref{17}%
), (\ref{18}).

\item[second:]  a ``reduction'' $\mathcal{R}$ of (normal) states on $%
\mathcal{M}$ to states on $\mathcal{H}$ such that for all states $\hat{\rho}$
on $\mathcal{H}$ it holds
\begin{equation}
\mathcal{R}\left( \left( \Gamma (T)U_{m}B_{n}^{*}\right) \mathcal{E}%
^{*}\left( \hat{\rho}\right) \left( \Gamma (T)U_{m}B_{n}^{*}\right)
^{*}\right) =V_{nm}\hat{\rho}V_{nm}^{*}\quad (n,m=1,\ldots ,N),  \label{52}
\end{equation}
where $(V_{nm})_{n,m=1}^{N}$ are unitary operators on $\mathcal{H}$.
\end{description}

That we can obtain as follows: We have already stated in section 2 that
\[
\left( |\mathrm{exp}\;(aK_{1}(g_{j}))-\mathrm{exp}\;(0)\rangle \right)
_{j=1}^{N}\quad (\emph{r}=1,2)
\]
are ONS in $\mathcal{M}$. We denote by $\mathcal{M}_{\emph{r}}$ $(\emph{r}%
=1,2)$ the corresponding $N$-- dimensional subspaces of $\mathcal{M}$. Then
for each $\emph{r}=1,2,$ there exists exactly one unitary operator $W_{\emph{%
r}}$ from $\mathcal{H}$ onto $\mathcal{M}_{\emph{r}}\subseteq \mathcal{M}$
with
\begin{equation}
W_{\emph{r}}|j\rangle =|\mathrm{exp}\;(aK_{\emph{r}}g_{j})-\mathrm{exp}\;(0)
\rangle \quad (j=1,\ldots ,N)  \label{53}
\end{equation}
We put
\begin{equation}
\mathcal{E}^{*}\left( \hat{\rho}\right) :=W_{1}\hat{\rho}W_{1}^{*} \Pi _{%
\mathcal{M}_{1}}\quad \left( \hat{\rho}\text{ state on }\mathcal{H}\right) ,
\label{54}
\end{equation}
where $\Pi _{\mathcal{M}_{\emph{r}}}$ denotes the projection onto $\mathcal{M%
}_{\emph{r}}$ $(\emph{r}=1,2)$. \newline
Describing the state $\hat{\rho}$ on $\mathcal{H}$ by
\begin{equation}
\hat{\rho}=\sum\limits_{s=1}^{N}\lambda _{s}|\hat{\Phi}_{s}\rangle \langle
\hat{\Phi}_{s}|  \label{55}
\end{equation}
with
\[
|\hat{\Phi}_{s}\rangle =\sum\limits_{j=1}^{N}c_{sj}|j\rangle ,
\]
where $\left( c_{sj}\right) _{s,j=1}^{N}$ fulfills (\ref{18}), we obtain
that $\rho =\mathcal{E}^{*}\left( \hat{\rho}\right) $ is given by (\ref{16})
and (\ref{17}). \newline
Now, for each state $\rho $ on $\mathcal{M}$ we put
\begin{equation}
\mathcal{R}(\rho ):=\frac{W_{2}^{*}\Pi _{\mathcal{M}_{2}}\rho W_{2}} {%
\mathrm{tr}_{\mathcal{M}}W_{2}^{*}\Pi _{\mathcal{M}_{2}}\rho W_{2}}
\label{56}
\end{equation}
Since
\[
\Pi _{\mathcal{M}_{2}}\Gamma (T)U_{m}B_{n}^{*}\mathcal{E}^{*} \left( \hat{%
\rho}\right) \left( \Gamma (T)U_{m}B_{n}^{*}\right) ^{*}= \Gamma
(t)U_{m}B_{n}^{*}\mathcal{E}^{*}\left( \hat{\rho}\right) \left(
\Gamma(T)U_{m}B_{n}^{*}\right) ^{*},
\]
we get
\[
\mathrm{tr}_{\mathcal{M}}W_{2}^{*}\Pi _{\mathcal{M}_{2}}\Gamma
(t)U_{m}B_{n}^{*}\mathcal{E}^{*}\left( \hat{\rho}\right) \left( \Gamma
(T)U_{m}B_{n}^{*}\right) ^{*}=1
\]
and
\[
\mathcal{R}\left( \Gamma (T)U_{m}B_{n}^{*}\mathcal{E}^{*} \left( \hat{\rho}%
\right) \left( \Gamma (T)U_{m}B_{n}^{*}\right) ^{*}\right) =
W_{2}^{*}\Gamma(T)U_{m}B_{n}^{*}W_{1}\hat{\rho}W_{1}^{*}\Pi _{\mathcal{M}%
_{1}} \left( \Gamma(T)U_{m}B_{n}^{*}\right) ^{*}W_{2}.
\]
As we have the equality
\[
\Pi _{\mathcal{M}_{1}}\left( \Gamma (T)U_{m}B_{n}^{*}
\right)^{*}W_{2}=\left( \Gamma (T)U_{m}B_{n}^{*}\right) ^{*}W_{2},
\]
which implies
\[
\mathcal{R}\left( \Gamma (T)U_{m}B_{n}^{*}\mathcal{E}^{*} \left( \hat{\rho}%
\right) \left( \Gamma (T)U_{m}B_{n}^{*}\right) ^{*}\right) =
W_{2}^{*}\Gamma(T)U_{m}B_{n}^{*}W_{1}\hat{\rho}W_{1}^{*} \left(
\Gamma(T)U_{m}B_{n}^{*}\right) ^{*}W_{2}
\]
Put
\begin{equation}
V_{nm}:=W_{2}^{*}\Gamma (T)V_{m}B_{n}^{*}W_{1}\quad (n,m=1,\ldots ,N),
\label{57}
\end{equation}
then $V_{nm}$ $(n,m=1,\ldots ,N)$ is an unitary operator on $\mathcal{H}$
and (\ref{52}) holds. One easily checks
\[
V_{nm}|j\rangle =\bar{b}_{nj}|j\otimes m\rangle \quad (j,m,n=1,\ldots ,N).
\]
Summarizing these, we have the following theorem:

\begin{ftheo}
\label{def22} Consider the channels on the set of states on $\mathcal{H}$
\begin{eqnarray}
\hat{\Lambda}_{nm} &:&=\mathcal{R}\circ \Lambda _{nm}\circ \mathcal{E}%
^{*}\quad (n,m=1,\ldots ,N)  \label{58} \\
\hat{\Theta}_{nm} &:&=\mathcal{R}\circ \Theta _{nm}\circ \mathcal{E}%
^{*}\quad (n,m=1,\ldots ,N)  \label{59}
\end{eqnarray}
where $\mathcal{R},$ $\mathcal{E}^{*},\Lambda _{nm},\Theta _{nm}$ are given
by (\ref{56}), (\ref{54}), (\ref{37}), (\ref{43}), respectively. Then for
all states $\hat{\rho}$ on $\mathcal{H}$, it holds
\begin{equation}
\hat{\Lambda}_{nm}\left( \hat{\rho}\right) =V_{nm}\hat{\rho}V_{nm}^{*}=\hat{%
\Theta}_{nm}\left( \hat{\rho}\right) \quad (n,m=1,\ldots ,N),  \label{60}
\end{equation}
where $V_{nm}\;(n,m=1,\ldots ,N)$ are the unitary operators on $\mathcal{H}$
given by (\ref{57}).
\end{ftheo}

\begin{frem}
\label{def23} Remember that the teleportation model according to $\left(
\Lambda _{nm}\right) _{n,m=1}^{N}$ works perfectly in the sense of the
remark \ref{def14}, and the model dealing with $\left( \Theta _{nm}\right)
_{n,m=1}^{N}$ was only asymptotically perfect for large $d$(i.e.,high
density or high energy of the beams). They can transfer to $\left( \hat{%
\Lambda}_{n,m}\right) $ , $\left( \hat{\Theta}_{nm}\right) $.
\end{frem}

\begin{fex}
\label{def24} We specialize
\[
\hat{K}_{1}h=\hat{K}_{2}h=\frac{1}{\sqrt{2}}\;h\quad \left( h\in L^{2}(%
\mathbf{R}^{k})\right) ,\text{ }\hat{T}=\mathbf{1.}
\]
Realizing the teleportation in this case means that Alice has to perform
measurements $\left( F_{nm}\right) $ in the whole space $\mathbf{R}^{k}$ and
also Bob (concerning $F_{+}$).
\end{fex}

\section{Alice and Bob are spatially separated}

We specialize the situation in section 4 as follows: We fix \vspace*{-3mm}

\begin{enumerate}
\item[-]  $t\in \mathbf{R}^{k}$

\item[-]  $X_{1},X_{2},X_{3}\subseteq \mathbf{R}^{k}$ are measurable
decomposition of $\mathbf{R}^{k}$ such that $l(X_{1})\neq 0$ and
\[
X_{2}=X_{1}+t:=\{x+t|\;x\in X_{1}\}.
\]
Put
\[
\begin{array}{llll}
\hat{T}h(x) & := & h(x-t)\qquad  & \left( x\in \mathbf{R}^{k}\,,\;h\in L^{2}(%
\mathbf{R}^{k})\right)  \\
\hat{K}_{\emph{r}}h & := & h\mathcal{X}_{X_{\emph{r}}} & \left( \emph{r}%
=1,2\,,\;h\in L^{2}(\mathbf{R}^{k})\right)
\end{array}
\]
\end{enumerate}

and assume that the function $f\in L^{2}\left( \mathbf{R}^{d}\right) $ has
the properties
\[
f\mathcal{X}_{X_{2}}=\hat{T}\left( f\mathcal{X}_{X_{1}}\right) ,\;f\mathcal{X%
}_{X_{3}}\equiv 0
\]
Then $\hat{T}$ is an unitary operator on $L^{2}\left( \mathbf{R}^{k}\right) $
and (\ref{48} ),(\ref{49}) hold. \newline
Using the assumption that $X_{1},X_{2},X_{3}$ is a measurable decomposition
of $\mathbf{R}^{k}$ we get immediately that
\[
G_{s}:=X_{s}\times \{1,\ldots ,N\}\quad (s=1,2,3)
\]
is a measurable decomposition of $G$. It follows that $\mathcal{M}=\mathcal{M%
}(G)$ is decomposed into the tensor product
\[
\mathcal{M}(G)=\mathcal{M}(G_{1})\otimes \mathcal{M}(G_{2})\otimes \mathcal{M%
}(G_{3}).
\]
\cite{FF1,FF2,FW}. According to this representation, the local algebras $%
\frak{A}(X_{s})$ corresponding to regions $X_{s}\subseteq \mathbf{R}^{d}$ $%
\left( s=1,2,3\right) $ are given by
\[
\begin{array}{llll}
\frak{A}(X_{1}) & := & \{A\otimes \mathbf{1}\otimes \mathbf{1};A\text{
bounded operator on }\mathcal{M}(G_{1})\}\qquad  &  \\
\frak{A}(X_{2}) & := & \{\mathbf{1}\otimes A\otimes \mathbf{1};A\text{
bounded operator on }\mathcal{M}(G_{2})\} &  \\
\frak{A}(X_{3}) & := & \{\mathbf{1}\otimes \mathbf{1}\otimes A;A\text{
bounded operator on }\mathcal{M}(G_{3})\} &
\end{array}
\]
One easily checks in our special case that
\[
F_{nm}\in \frak{A}(X_{1})\otimes \frak{A}(X_{2})\quad (n,m=1,\ldots ,N)
\]
and $\mathcal{E}^{*}\left( \hat{\rho}\right) $ gives a state on $\frak{A}%
(X_{1})$ (the number of particles outside of $G_{1}$ is $0$ with probabiliy $%
1$ ). That is, Alice has to perform only local measurements inside of the
region $X_{1}$ in order to realize the teleportation processes described in
section 4 or measure the state $\mathcal{E}^{*}\left( \hat{\rho}\right) $.
On the other hand, $\Lambda _{nm}\left( \mathcal{E}^{*}\left( \hat{\rho}%
\right) \right) $ and $\Theta _{nm}\left( \mathcal{E}^{*}\left( \hat{\rho}%
\right) \right) $ give local states on $\frak{A}(X_{2})$ such that by
measuring these states Bob has to perform only local measurements inside of
the region $X_{2}$. The only problem could be that according to the
definition (\ref{43}) of the channels $\Theta _{nm}$ Bob has to perform the
measurement by $F_{+}$ which is not local. However, as we have already
stated in the remark \ref{def20}, this problem can be avoided if we replace $%
F_{+}$ by $F_{+,X_{2}}\in \frak{A}(X_{2})$. \newline
Therefore we can describe the special teleportation process as follows: We
have a beam being in the pure state $|\eta \rangle \langle \eta |$ (\ref{40}%
) . After splitting, one part of the beam is located in the region $X_{1}$
or will go to $X_{1}$ (cf. remark \ref{def10}) and the other part is located
in the region $X_{2}$ or will go to $X_{2}$. Further, there is a state $%
\mathcal{E}^{*}\left( \hat{\rho}\right) $ localized in the region $X_{1}$.
Now Alice will perform the local measurement inside of $X_{1}$ according to $%
F=\sum\limits_{n,m}z_{nm}F_{nm}$ involving the first part of the beam and
the state $\mathcal{E}^{*}(\rho )$. This leads to a preparation of the
second part of the beam located in the region $X_{2}$ which can be
controlled by Bob, and the second part of the beam will show the behaviour
of the state $\Lambda _{nm}\left( \mathcal{E}^{*}\left( \hat{\rho}\right)
\right) =\Theta _{nm}\left( \mathcal{E}^{*}\left( \hat{\rho}\right) \right) $
if Alice's measurement shows the value $z_{nm}$. Thus we have teleported the
state $\hat{\rho}$ on $\mathcal{H}$ from the region $X_{1}$ into the region $%
X_{2}$.

\end{document}